\documentclass{iau} 
\usepackage{graphicx,natbib,bm,color}

\newcommand{\EQ}{\begin{equation}}
\newcommand{\EN}{\end{equation}}
\newcommand{\EQA}{\begin{eqnarray}}
\newcommand{\ENA}{\end{eqnarray}}

\newcommand{\Eq}[1]{Equation~(\ref{#1})}

\newcommand{\Fig}[1]{Figure~\ref{#1}}

\newcommand{\Tab}[1]{Table~\ref{#1}}

\newcommand{\bra}[1]{\langle #1\rangle}

{}
{}
{}

{}
{}
{}
{}
{}
{}
{}
{}
{}
{}
{}
{}
{}
{}
{}
{}
{}
{}
{}

{}
{}
{}

{}

{}
{}

\newcommand{\zzz}{\hat{\mbox{\boldmath $z$}} {}}

\newcommand{\kk}{\bm{k}}

\newcommand{\xx}{\bm{x}}

\newcommand{\uu}{\mbox{\boldmath $u$} {}}

\newcommand{\bb}{\mbox{\boldmath $b$} {}}

\newcommand{\nab}{\mbox{\boldmath $\nabla$} {}}
\newcommand{\OO}{\bm{\Omega}}

\newcommand{\ii}{{\rm i}}

\newcommand{\diag}{{\rm diag}  \, {}}

\newcommand{\dd}{{\rm d} {}}

\def\degr{\hbox{$^\circ$}}
\def\la{\mathrel{\mathchoice {\vcenter{\offinterlineskip\halign{\hfil
$\displaystyle##$\hfil\cr<\cr\sim\cr}}}
{\vcenter{\offinterlineskip\halign{\hfil$\textstyle##$\hfil\cr<\cr\sim\cr}}}
{\vcenter{\offinterlineskip\halign{\hfil$\scriptstyle##$\hfil\cr<\cr\sim\cr}}}
{\vcenter{\offinterlineskip\halign{\hfil$\scriptscriptstyle##$\hfil\cr<\cr\sim\cr}}}}}
\def\ga{\mathrel{\mathchoice {\vcenter{\offinterlineskip\halign{\hfil
$\displaystyle##$\hfil\cr>\cr\sim\cr}}}
{\vcenter{\offinterlineskip\halign{\hfil$\textstyle##$\hfil\cr>\cr\sim\cr}}}
{\vcenter{\offinterlineskip\halign{\hfil$\scriptstyle##$\hfil\cr>\cr\sim\cr}}}
{\vcenter{\offinterlineskip\halign{\hfil$\scriptscriptstyle##$\hfil\cr>\cr\sim\cr}}}}}
\def\Co{\mbox{\rm Co}}
\def\Ro{\mbox{\rm Ro}}

\def\Co{\mbox{\rm Co}}

\def\kf{k_{\rm f}}

\def\Brms{B_{\rm rms}}

\def\urms{u_{\rm rms}}

\def\Beq{B_{\rm eq}}

\newcommand{\yr}{\,{\rm yr}}

\newcommand{\Gyr}{\,{\rm Gyr}}

\newcommand{\yjmp}[3]{ #1, \textit{JMP,} \textit{#2}, #3}
\newcommand{\yapj}[3]{ #1, \textit{ApJ,} \textit{#2}, #3}
\newcommand{\yarep}[3]{ #1, \textit{Astron.\ Rep.,} \textit{#2}, #3}

\newcommand{\yapjl}[3]{ #1, \textit{ApJ,} \textit{#2}, #3}

\newcommand{\yan}[3]{ #1, \textit{Astron.\ Nachr.,} \textit{#2}, #3}

\newcommand{\yana}[3]{ #1, \textit{A\&A,} \textit{#2}, #3}

\newcommand{\yass}[3]{ #1, \textit{Ap\&SS,} \textit{#2}, #3}
\newcommand{\ygafd}[3]{ #1, \textit{Geophys.\ Astrophys.\ Fluid Dyn.,} \textit{#2}, #3}

\newcommand{\yaraa}[3]{ #1, \textit{ARA\&A,} \textit{#2}, #3}

\newcommand{\yprl}[3]{ #1, \textit{Phys.\ Rev.\ Lett.,} \textit{#2}, #3}

\newcommand{\ymn}[3]{ #1, \textit{MNRAS,} \textit{#2}, #3}
\newcommand{\ynat}[3]{ #1, \textit{Nature,} \textit{#2}, #3}
\newcommand{\yptrsa}[3]{ #1, \textit{Phil. Trans. Roy. Soc. London A,} \textit{#2}, #3}
\newcommand{\ysci}[3]{ #1, \textit{Science,} \textit{#2}, #3}
\newcommand{\ysph}[3]{ #1, \textit{Solar Phys.,} \textit{#2}, #3}

\newcommand{\yprd}[3]{ #1, \textit{Phys.\ Rev.\ D,} \textit{#2}, #3}
\newcommand{\ypre}[3]{ #1, \textit{Phys.\ Rev.\ E,} \textit{#2}, #3}

\newcommand{\yjour}[4]{ #1, \textit{#2}, \textit{#3}, #4}
\newcommand{\djour}[3]{ #1, \textit{#2}, doi:#3}

\newcommand{\ybook}[3]{ #1, \textit{#2} (#3)}
\newcommand{\yproc}[5]{ #1, in \textit{#3}, ed.\ #4 (#5), #2}

\newcommand{\sapj}[1]{ #1, \textit{ApJ}, submitted}
\newcommand{\sana}[2]{ #1, \textit{A\&A}, submitted, arXiv:#2}

\title
{Magnetic field evolution in solar-type stars}

\author[Axel Brandenburg]   
{Axel Brandenburg$^{1,2,3,4,5}$}

\affiliation{
$^1$Nordita, KTH Royal Institute of Technology and Stockholm University,\\
Roslagstullsbacken 23, SE-10691 Stockholm, Sweden;
email: {\tt brandenb@nordita.org} \\[\affilskip]
$^2$Department of Astronomy, Stockholm University, SE-10691 Stockholm, Sweden\\
$^3$JILA and Laboratory for Atmospheric and Space Physics, Univ. Colorado, Boulder, USA\\
$^4$McWilliams Center for Cosmology, Carnegie Mellon University, Pittsburgh, PA 15213, USA\\
$^5$Faculty of Natural Sciences and Medicine, Ilia State University, 0194 Tbilisi, Georgia
}

\pubyear{2020}
\volume{354}  
\jname{Solar and Stellar Magnetic Fields: Origins and Manifestations}
\editors{A. G. Kosovichev, K. G. Strassmeier \& M. Jardine, eds.\\
\today,~ $ $Revision: 1.34 $ $}
\begin{document}

\maketitle

\begin{abstract}
We discuss selected aspects regarding the magnetic field evolution of
solar-type stars.
Most of the stars with activity cycles are in the range where the
normalized chromospheric Calcium emission increases linearly with the
inverse Rossby number.
For Rossby numbers below about a quarter of the solar value,
the activity saturates and no cycles have been found.
For Rossby numbers above the solar value, again no activity cycles have
been found, but now the activity goes up again for a major fraction of
the stars.
Rapidly rotating stars show nonaxisymmetric large-scale magnetic fields,
but there is disagreement between models and observations regarding the
actual value of the Rossby number where this happens.
We also discuss the prospects of detecting the sign of magnetic
helicity using various linear polarization techniques both at the
stellar surface using the parity-odd contribution to linear polarization
and above the surface using Faraday rotation.
\end{abstract}

\firstsection 
\section{Introduction}

The purpose of this paper is to discuss recent results relevant for
understanding the connection between observations and simulations
of magnetic fields in solar-like stars.
We focus on observational measures of stellar activity, the occurrence
of activity cycles in other stars, and new ways of interpreting stellar surface
magnetic fields in terms of linear polarization measurements.
We also address the possibility of a radial sign reversal of the star's
magnetic helicity some distance above the stellar surface.

The Sun's magnetic field exhibits remarkable regularity in space and time,
as is demonstrated by Maunder's butterfly diagram \citep{Mau04}.
Although comparable diagrams are only now beginning to become possible
for other stars \citep[see, e.g.,][]{Alvarado}, there is clear evidence
that cyclic chromospheric variability is ubiquitous.
This became clear after \cite{Wil63,Wil78} selected a set of stars
that were then monitored for the next three decades at Mount Wilson
\citep{Bal+95}.

Much of our knowledge on stellar magnetic activity comes from
understanding the Sun's magnetic field.
The occurrence of a fairly regular 11-year activity cycle is of course
one of its main characteristics.
A cycle as clear as that of the Sun has not been observed for any of
the other stars monitored so far.
The perhaps best observed cycle is that of HD 81809, but that star is a binary
and the cyclic component is not a main sequence star, but a subgiant;
see \cite{Ege18} for a detailed discussion of this interpretation.
The lack of equally clear cycles makes one wonder whether the Sun is
perhaps a special case.
Other evidence in favor of such thinking is the fact that in a diagram
of cycle period versus rotation period, the Sun lies right between two
different branches, the high and low activity branches \citep{BV07}.
There are a few other aspects suggesting that the Sun is special.
Some of them are discussed below.

In the earlier work of \cite{BST98}, which also showed two distinct
activity branches, the Sun appeared closer to the inactive branch
and not really between two branches.
One reason why the Sun was closer to the inactive branch was the fact that
they used the 10 year cycle period obtained by \cite{Bal+95}
for the time interval for which their solar $\bra{R'_{\rm HK}}$ data
were determined from nightly moonlight observations.
Using instead the 11 year cycle period determined for the full record
since the time of \cite{Schwabe1844} and before \citep[since the time of
the Maunder minimum, as was established by][]{Edd76} yields a position of
the Sun that is now further away from the inactive branch \citep{BMM17},
although it is still not really between the two branches.
Regarding the clear cyclicity,
it should also be kept in mind that the Sun is relatively old compared
with many of the stars for which cycles have been obtained.
It is therefore possible that there is some sort of selection effect
for why only the Sun has such a well defined cycle.

We begin by discussing first the relation between rotation period and
stellar activity.
Also this relation suggests that the Sun's location in that diagram
is between two different modes of behavior, which is when the activity
attains a minimum as a function of rotation periods.
Both for faster and for slower rotation, the activity increases relative
to that of the Sun, at least approximately; see the work \cite{BG18},
who offered an interpretation in terms of the stellar differential
rotation changing from solar-like to antisolar-like differential rotation
right at the Sun's rotation rate.
So, again, the Sun appears to take a special position among the
many other stars.
We finish by discussing new ideas for determining solar and
stellar magnetic helicity from linear polarization measurements.
This technique, however, has so far only been applied to the Sun.

\section{Activity versus rotation}

Stellar variability is usually characterized by the chromospheric
Ca~{\sc ii}~H+K line emission, normalized by the bolometric flux
and corrected for photospheric contributions to give $R'_{\rm HK}$.
This quantity is believed to be a good measure of the mean magnetic
field normalized by the equipartition field strength, which is defined
based on the kinetic energy as $B_{\rm eq}=\sqrt{4\pi\rho}\urms$,
where $\urms$ is the rms value of the turbulent velocity and $\rho$
is the local gas density.
\cite{Sch89} estimated that $R'_{\rm HK}$ is related to the rms magnetic
field strength, $\Brms$, through
\EQ
R'_{\rm HK}\propto (\Brms/\Beq)^{0.5},
\EN
where $\Brms$ is expressed in
terms of a filling factor $f$ and the typical field strength in spots
$B_{\rm spot}$, which can be estimated by the photospheric pressure
$p_{\rm phot}$ as $B_{\rm spot}=\sqrt{8\pi p_{\rm phot}}$.
\cite{SL85} thus proposed
\EQ
\Brms=f\,B_{\rm spot}.
\EN
As stars become more active, $f$ increases up to the point when the
entire stellar surface is covered with spots, leading to saturation
as $f$ cannot increase beyond unity ($f\le1$ by definition).

Stellar rotation with angular velocity $\Omega$ affects the dynamo
process through the Coriolis force, $2\OO\times\uu$, and its strength
is characterized by the Coriolis number, $\Co=2\Omega\tau$, where $\tau$
is some measure of the turnover time.
One often quotes the Rossby number, $\Ro=P_{\rm rot}/\tau$, where
$P_{\rm rot}=2\pi/\Omega$ is the rotation period.
The two parameters are then related to each other through $\Ro=4\pi/\Co$,
although \cite{BST98} defined the Rossby number simply as $\Co^{-1}$.
There are also other reasons why a statement about the Rossby number
should be taken with care.

To know the actual values of $\Ro$ or $\Co$, one has to agree on a good
definition for $\tau$.
From an observational point of view, all that matters is that $\tau$ is
a monotonically decreasing function of stellar mass, or, in practice,
a monotonically increasing function of the color $B-V$.
But even then there can be significant uncertainties.
In the relevant color range of $B-V$ between 0.6 and 0.75,
the turnover times of \cite{BK10} are about 2.4 times longer than
those of \cite{Noy+84}.
In the work of \cite{BG18}, for example, the turnover times of
\cite{Noy+84} were adopted.
For definitiveness, we refer to those times as $\tau_{\rm Noy}$.
The original calculation of $\tau_{\rm Noy}$ was based upon stellar
mixing length models of \cite{Gil80}, where $\tau_{\rm Noy}=\ell/\urms$
was based on the local mixing length $\ell$ and the rms velocity $\urms$
about one scale height above the bottom of the convection zone.

\begin{figure}[b]
\begin{center}
\includegraphics[width=\textwidth]{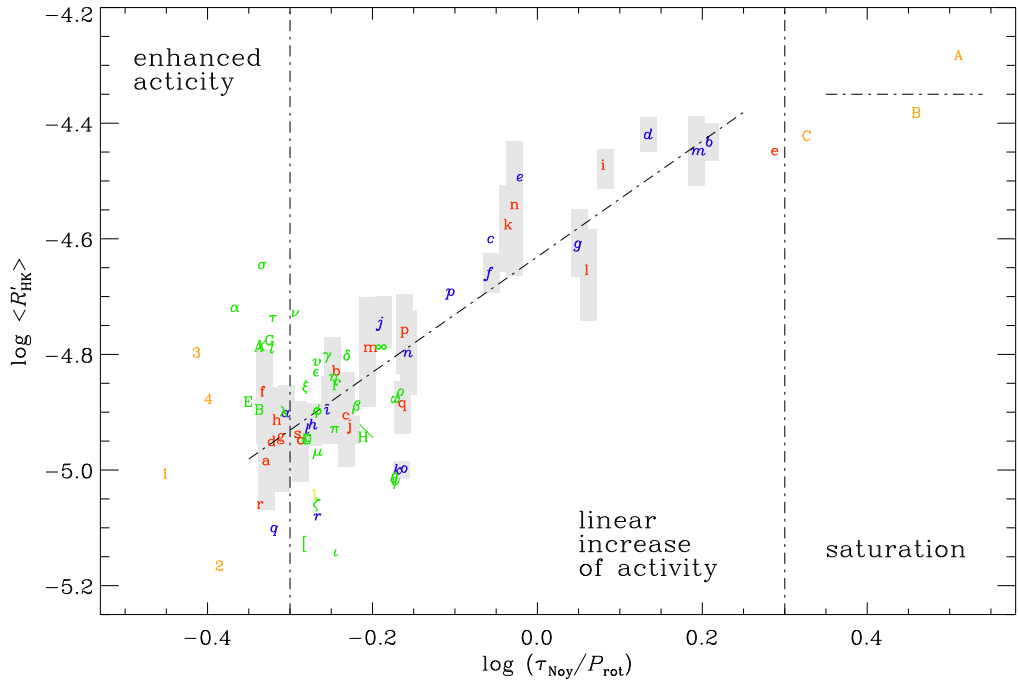} 
\caption{
Activity versus Rossby number, showing saturation for
$\log(\tau_{\rm Noy}/P_{\rm rot})>0.3$ (see \Tab{TSum1})
and enhanced activity for many stars in the range
$\log(\tau_{\rm Noy}/P_{\rm rot})<-0.3$ (see also \Tab{TSum2}).
}\label{pRHK_4Gyr_all_revised_IAU}
\end{center}
\end{figure}

It has been known for some time that $R'_{\rm HK}$ scales with $\Co$
\citep{Noy+84,Vil84} and is therefore inversely proportional to $\Ro$,
until saturation is reached for rapid rotation.
As alluded to above, saturation is often interpreted in terms of the
filling factor of the magnetic field on the stellar surface approaching
unity.
However, none of the stars with cycles are anywhere near saturation,
so they obey $\bra{R'_{\rm HK}}\propto\Co$, but there is some leveling
off for non-cyclic stars with $\log(\tau_{\rm Noy}/P_{\rm rot})>0.3$.
This is shown in \Fig{pRHK_4Gyr_all_revised_IAU}, where we have combined
data from the three stars of \Tab{TSum1} (see the orange labels A--C
for HD~17925, 131156A, and 131156B, respectively) with data from
\cite{BMM17}, red characters for K dwarfs and blue ones for G and
F dwarfs, and \cite{BG18}, green symbols, for the stars of the open
cluster M~67 with rotation periods estimated from gyrochronology based
on an estimated age of $4\Gyr$; see \cite{BG18} for details.
The orange numbers 1--4 for small values of $\log(\tau/P_{\rm rot})$
are for other field stars discussed already in \cite{BG18}; see also
\Tab{TSum2}.
For those stars, we have computed their ages from Equations~(12)--(14)
of \cite{MH08} as
\EQ
t = \left\{P_{\rm rot}/ [0.407\,(B-V-0.495)^{0.325}]\right\}^{1.767},
\label{GyroChron}
\EN
provided $B-V>0.495$; see also \cite{BMM17}.

Some of the stars of M67 show increasing activity with decreasing
$\tau_{\rm Noy}/P_{\rm rot}$ or decreasing rotation rate, which suggests
that the activity rises again as they slow down further.
This has been interpreted by \cite{BG18} as possible evidence for
anti-solar differential rotation, which is theoretically expected for
very slow rotation \citep{Gil77}.
The absolute differential rotation in the regime of antisolar differential
rotation is known to be stronger than that in the regime of solar-like
differential rotation \citep{GYMRW14,KKB14}, and also leads to larger
magnetic activity for stars with smaller $\tau_{\rm Noy}/P_{\rm rot}$
\citep{KKKBOP15}, thus explaining their enhanced activity.

\begin{table}[t!]\caption{
Parameters of three active stars in the saturation range.
The data are taken from \cite{Noy+84} and the gyrochronological age is
computed from the formula of \cite{MH08}; see also \Eq{GyroChron}.
}\vspace{12pt}\centerline{\begin{tabular}{ccccccc}
Label & HD & $B$--$V$ & $\tau\;$ [d] & $P_{\rm rot}$ [d] &
$\log\bra{R'_{\rm HK}}$ & Age [Gyr] \\
\hline
A &  17925 & 0.87 & 21.3 & ~6.6 & $-4.28$ & 0.2 \\ 
B &131156A & 0.76 & 17.8 & ~6.2 & $-4.38$ & 0.3 \\ 
C &131156B & 1.17 & 24.3 & 11.5 & $-4.42$ & 0.5 \\ 
\label{TSum1}\end{tabular}}
\end{table}

\begin{table}[t!]\caption{
Stars in the range of enhanced activity.
HD~187013 and HD~224930 do indeed show enhanced activity.
For HD~187013, $B-V$ is outside the range of applicability
of \Eq{GyroChron}.
}\vspace{12pt}\centerline{\begin{tabular}{ccccccc}
Label & HD & $B$--$V$ & $\tau\;$ [d] & $P_{\rm rot}$ [d] &
$\log\bra{R'_{\rm HK}}$ & Age [Gyr] \\
\hline
1 & 141004 & 0.60 &  9.1 & 25.8 & $-5.00$ & 5.6 \\ 
2 & 161239 & 0.65 & 12.0 & 29.2 & $-5.16$ & 5.5 \\ 
3 & 187013 & 0.47 &  3.1 & ~8.0 & $-4.79$ & old \\ 
4 & 224930 & 0.67 & 13.1 & 33.0 & $-4.88$ & 6.4 \\ 
\label{TSum2}\end{tabular}}
\end{table}

However, when comparing with simulations, the situation is still somewhat
puzzling.
In fact, numerical simulations of convectively driven dynamo action
show that there is another transition, where the large-scale magnetic
field becomes predominately nonaxisymmetric with an $m=1$ azimuthal
modulation \citep{Viviani18}.
Present models show that these two transitions happen more or less
at the same value of $\Co$, but observations shows that the two transitions
happen at different values: anti-solar differential rotation for $\Ro\ga2.0$
\citep{BG18} and nonaxisymmetric large-scale field for $\Ro\la0.5...1.0$;
see Table~5 of \cite{Viviani18}, who refer to data of \cite{Lehtinen16}.
In fact, recent work of \cite{Lehtinen20} suggests that the transition
from nonaxisymmetric to axisymmetric magnetic fields might be
accompanied a change in the slope of the chromospheric activity versus
Rossby number diagram, which becomes particularly clear when they combine
data of main sequence stars with those of subgiants and giants.
The lack of corresponding features in data from numerical simulations
is an important shortcoming of current simulations.
It could be that some sort of ``renormalization'' is required when
comparing numerical simulations with observational data.
Indeed, simulations are long known to yield cyclic behavior only for
rotation rates that exceed the solar value by about a factor of three
\citep{BMBBT11}.

Not all stars necessarily slow down with age.
Their Sun-like activity cycles may just disappear, but they
would still spin rapidly \citep{MvS17}.
If their magnetic field topology develops predominantly small scales,
as has now been demonstrated by \cite{Metcalfe19}, magnetic breaking
would become progressively inefficient.
This idea emerged when discrepancies between helioseismic ages and
gyrochronological ages became apparent \citep{vanSaders2016}.
\cite{BG18} speculated that this could still be reconciled with the
possibility of antisolar differential rotation if there is a bifurcation
into two possible scenarios: stars that make the transition to antisolar
differential rotation as a result of a sufficiently chaotic evolution, and
others that just change their field topology and remain rapidly spinning.
Demonstrating this with actual models would clearly be a next important step.

\section{Cycle frequency versus activity}

A systematic dependence of cycle frequency $\omega_{\rm cyc}$
($=2\pi/P_{\rm cyc}$) on rotation rate $\Omega$ was first found
by \cite{NWV84} based on the early analyses of the sample of
\cite{Wil63,Wil78} measured at Mount Wilson.
They found $\omega_{\rm cyc}\propto\Omega^{1.25}$.
It is important to emphasize that {\em the exponent is larger than unity},
which has long been a theoretical difficulty to explain.
An early theoretical analysis by \cite{KRS83} based on
the fastest growing linear eigenmode yielded promising results with
$\omega_{\rm cyc}\propto\Omega^{4/3}$, but very different solutions
were obtained for nonlinear saturated dynamos \citep{Tob98}.
This was also emphasized by \cite{BST98}, who proposed that spatial
nonlocality could be strong enough so that only solutions with the lowest
wavenumber would exist.
This seems to be the best explanation even today.

\begin{table}[t!]\caption{
Selected stars with well defined cycles using data from \cite{BMM17}.
}\vspace{12pt}\centerline{\begin{tabular}{ccccccccc}
Label & HD & $B$--$V$ & $\tau\;$ [d] & $P_{\rm rot}$ [d] &
$\log\bra{R'_{\rm HK}}$ & Age [Gyr] & $P_{\rm cyc}$ [yr] & comp \\
\hline
a (blue) &     Sun&0.66&12.6&$25.4$ &$-4.90$&4.6&$11.0$&~6.6 \\
c (blue) & ~~10476&0.84&20.6&$35.2$ &$-4.91$&4.9&$~9.6$&~9.3 \\
f (blue) & ~~26965&0.82&20.1&$43.0$ &$-4.87$&7.2&$10.1$&10.9 \\
m (blue) &  160346&0.96&22.7&$36.4$ &$-4.79$&4.4&$~7.0$&~8.5 \\
\label{TSum3}\end{tabular}}
\end{table}

As already mentioned in the introduction, there are only very few stars
that show well defined cycles that are nearly as clean as that of the Sun.
We have listed the properties of these stars in \Tab{TSum3}, where
we also list the cycle periods, $P_{\rm cyc}$, as well as those
{\em computed} from the formula of \cite{BMM17} that assumes that
the stars lie exactly on the long-period branch.
The stars cover the full range in $\log(\tau_{\rm Noy}/P_{\rm rot})$
from $-0.3$ to $0.2$; see \Fig{pRHK_4Gyr_all_revised_IAU}.
The ages of those stars are in the range from $4.4$ to $4.9\Gyr$,
except for HD~26965, which is $7.2\Gyr$.
This shows that all stars with well defined cycles are old stars.

Furthermore, the analysis of many cycle data by \cite{Bal+95}
suggested the existence of multiple cycles.
Their reality remains debated even today \citep{BSaikia16,OLKP18}.
The work of \cite{BST98} and \cite{BMM17} suggested two nearly parallel
branches of values of $\omega_{\rm cyc}/\Omega$ versus $\bra{R'_{\rm HK}}$.
The lower branch has cycle periods that are about six times longer than
those on the regular (upper) branch, where also the Sun was thought to
be located, if we adopted the $10\yr$ period; see the discussion above.
The two branches were originally called active and inactive branches,
because they were also well separated with respect to the vertical line
$\lg\bra{R'_{\rm HK}}=-4.65$.
\cite{BMM17} suggested, however, that (i) the branches are now well
overlapping and that (ii) stars younger than $3\Gyr$ might exhibit both
shorter and longer cycles simultaneously.
It should be remembered that these stars tend to be rapid rotators,
whose large-scale magnetic field is expected to be nonaxisymmetric
\citep{Viviani18}.
Such a magnetic field is similar to that of a dipole lying in the
equatorial plane and with opposite polarities at longitudes that are
$180\degr$ away from each other.


Nonaxisymmetric magnetic fields have long been predicted based on
mean-field models with an anisotropic $\alpha$ effect, and a tensor
$\alpha_{ij}$ whose diagonal components do not all have the same value.
Since the early work of \cite{Rue78}, it was known that at rapid
rotation, $\alpha_{ij}$ attains an additional piece proportional
to $\Omega_i\Omega_j$, so the tensor approaches the form
\EQ
\alpha_{ij} \to \alpha_0 \Big(\delta_{ij}-\Omega_i\Omega_j/\Omega^2\Big),
\label{AlphaTensor}
\EN
showing that the component of $\alpha$ in the $\OO$ direction vanishes.
In other words, if the $\OO$ direction corresponds to the $z$ direction
in Cartesian coordinates, $\alpha_{ij}$ is proportional to
$\diag(\alpha_{xx},\,\alpha_{yy},\,0)$.
In general, there can also be off-diagonal components, but those are
not important for the present discussion.
The main point is that with $\alpha_{zz}=0$, the dynamo-generated
magnetic field is, in Cartesian geometry, always horizontal.
In a sphere, it then corresponds to a dipole lying in the equatorial
plane; see Fig.~3(a) of \cite{MB95}.

A simple example of such a field is that generated by the Roberts
flow~I \citep[the first of four flows I--IV studied by][]{Rob72}.
This flow takes the form
\EQ
\uu=\nab\times\psi\zzz+\kf\psi\zzz,
\EN
which is a prototype example for modeling magnetic fields generated by
an $\alpha$ effect.
Here, the $\alpha$ tensor is indeed of the form of \Eq{AlphaTensor}
with $\alpha_{ij}=\diag(\alpha_0,\,\alpha_0,\,0)$ and some coefficient
$\alpha_0$.
It is also a model of the magnetic field generated in the Karlsruhe
dynamo experiment, which, in turn, is an idealized model of the geodynamo
\citep{RRAF02,RB03}.
The Coriolis number of the geodynamo is extremely large---much larger
than that of any of the observed stars.
It therefore tends to exaggerate the effects of rotation.

Observationally, nonaxisymmetric magnetic fields have been inferred
from light curve modeling and Doppler imaging
\citep[see, e.g.,][]{Kochukhov}.
However, we also know that stars with nonaxisymmetric magnetic field
can exhibit what is known as flip-flop phenomenon \citep{Jetsu}.
This means that the two opposite polarities alternate in strengths in
a cyclic fashion.

It is generally expected that such variations correspond to a mixed
parity solution of the type originally investigated by \cite{RWBMT90}.
Subsequent work of \cite{MBBT95} found it difficult to obtain such
solutions with their more realistic simulations.
Qualitative discussions have also been offered by \cite{EK05}.

Another related question concerns the cycle period observed for stars
on the active or long-period branch discussed above.
\cite{Gue19a} suggest that some sort of magnetic shear instability
might be responsible for cycle periods comparable to those of the Sun.
In this connection, there is also the question whether the
Tayler instability might play a role; see \cite{Gue19b}.
An important question concerns the surface appearance of magnetic fields
from the two branches of short and long cycle periods, especially for
young and rapidly rotating stars.
Are they really nonaxisymmetric and how can we understand the observed
occurrence of multiple cycles, i.e., the occurrence of multiple branches
with the same stars on both of them?
Modeling this convincingly would be a major step forward in understanding
the truth behind these two branches.

\section{New twists to polarimetric measurements}

Zeeman Doppler Imaging (ZDI) provides a powerful tool for characterizing
the actual magnetic field structure and its temporary changes.
Both in solar and stellar physics, one tends to display the results
directly in terms of the full magnetic field vector.
Those results are in general subject to the $180\degr$ ambiguity,
which is also sometimes referred to as the $\pi$ ambiguity.
In the solar context, this $\pi$ ambiguity might be an
important source of error in calculating the sign of the Sun's magnetic
helicity at large length scales.
It may therefore be advantageous to work directly with the Stokes
parameters \citep{BBKMRPS19,Bra19,Pra+20}.

Magnetic helicity is a quantity that characterizes the handedness of
the magnetic field.
Its sign would change if one looked at the star through a mirror.
In this connection, it is important to realize that the Stokes $Q$
and $U$ parameters can directly be expressed in terms of a quantity that
characterizes the sense of handedness.
This technique is routinely employed in cosmology, where one expresses
$Q$ and $U$ in terms of what is known as the parity even $E$ and the
parity odd $B$ polarizations.
To obtain $E$ and $B$, one expands $Q$ and $U$ not in terms of
the ordinary spherical harmonics, but in terms of spin-2 spherical
harmonics, $_2Y_{\ell m}(\theta,\phi)$; see \cite{Goldberg67}.
One then obtains $E$ and $B$ as the real and imaginary parts of the
transformed quantity in the form \citep{Kamion97,SZ97,ZS97,Dur08,KK16}
\EQ
E+\ii B\equiv R=\sum_{\ell=2}^{N_\ell}\sum_{m=-\ell}^{\ell}
\tilde{R}_{\ell m} Y_{\ell m}(\theta,\phi),
\EN
where the $\tilde{R}_{\ell m}$ are given by
\EQ
\tilde{R}_{\ell m}=\int_{4\pi}
(Q+\ii U)\,_2 Y_{\ell m}^\ast(\theta,\phi)\,
\sin\theta\,\dd\theta\,\dd\phi.
\label{EBfromQU}
\EN
In spectral space, we then define
$\tilde{E}_{\ell m}=(\tilde{R}_{\ell m}+\tilde{R}_{\ell,\,-m}^\ast)/2$
as the parity-even part and
$\tilde{B}_{\ell m}=(\tilde{R}_{\ell m}-\tilde{R}_{\ell,\,-m}^\ast)/2\ii$
as the parity-odd part, where the asterisk means complex conjugation.
This has recently been done for the Sun's magnetic field using synoptic
vector magnetograms, which yield a global map.
It turns out that $E$ is indeed even about the equator and $B$ is odd
about the equator.
Therefore, $\tilde{E}_{\ell m}$ has contributions mainly from even values
of $\ell$ and $\tilde{B}_{\ell m}$ has mainly contributions from odd
values of $\ell$.
As a useful proxy, one can therefore employ the correlators
\EQ
K_\ell^+=\tilde{E}_\ell\tilde{B}_{\ell+1}^\ast\quad\mbox{and}\quad
K_\ell^-=\tilde{E}_\ell\tilde{B}_{\ell-1}^\ast,
\label{Kdef1}
\EN
respectively, for different values of $\ell$.
There have also been attempts to determine the handedness of
magnetic fields in the solar neighborhood of the interstellar medium
\citep{Bracco19}.

To illustrate the decomposition further, we now discuss the
corresponding Cartesian decomposition.
It reads \citep{Dur08}
\EQ
\tilde{R}(k_x,k_y)=-(\hat{k}_x-\ii \hat{k}_y)^2\tilde{P}(k_x,k_y),
\label{Rkxky}
\EN
where
\EQ
\tilde{P}(k_x,k_y)=\int P(x,y) e^{-\ii\kk\cdot\xx} \dd^2x
\EN
is the Fourier transform of $P=Q+\ii U$, and $\kk=(k_x,k_y)$
$\xx=(x,y)$ are two-dimensional vectors.
The return transform, here written for $R$, is given by
\EQ
R(x,y)=\int \tilde{R}(k_x,k_y) e^{\ii\kk\cdot\xx} \dd^2x/(2\pi)^2.
\EN

\begin{figure}[t!]
\begin{center}
\includegraphics[width=\textwidth]{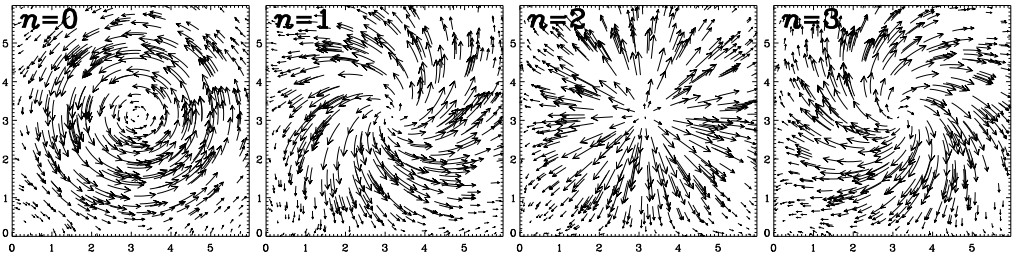}
\caption{
Plots of $e^{\ii\pi n/4}(k_x+\ii k_y)/k^3$ for $n=0$, 1, 2, and 3.
Different values of the phase sample $E$ patterns ($n=0$ and $2$)
and $B$ patterns ($n=1$ and $3$) in a continuous fashion.
}\label{ppp1}
\end{center}
\end{figure}

\begin{figure}[t!]
\begin{center}
\includegraphics[width=\textwidth]{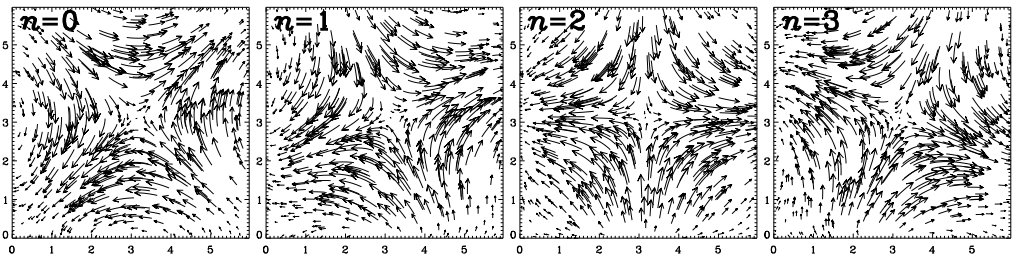}
\caption{
Plots of $e^{\ii\pi n/4}(k_x-\ii k_y)/k^3$ for $n=0$, 1, 2, and 3.
Unlike the case with $(k_x+\ii k_y)$, shown in \Fig{ppp1}, these
patterns do not correspond to the usual $E$ or $B$ polarizations.
}\label{ppp2}
\end{center}
\end{figure}

To generate a two-dimensional vector field, it suffices to 
combine the vector $\bb=(b_x,b_y)$ into a single complex field,
${\cal B}(x,y)=b_x+\ii b_y$.
Next, to generate a periodic pattern, we use the complex wavenumber
${\cal K}\equiv k_x+\ii k_y$ and generate a pattern in Fourier space
as the simplest possible nontrivial analytic function,
$\tilde{\cal B}({\cal K})={\cal K}$, along with some phase factor
$e^{\ii\phi}$ with phase $\phi$; see \Fig{ppp1}.
We also adopt its complex conjugate, ${\cal K}$; see \Fig{ppp2}.
In both cases we normalize by $|{\cal K}|^3$ to reduce the values
further away from the origin.

We see that the different patterns of
$\tilde{\cal B}({\cal K})=e^{\ii\phi}{\cal K}$
correspond to qualitatively different patterns.
For $n=0$ and 2, we obtain parity-even patterns ($E$ polarizations with
different signs of $E$), whereas for $n=1$ and 3, we obtain parity-odd
patterns ($B$ polarizations with different signs of $B$).
For $e^{\ii\phi}{\cal K}^*$, on the other hand, the patterns are quite
different, and they correspond to just the inverse of the former
ones in the sense that if one multiplies the $\tilde{\cal B}({\cal K})$
of \Fig{ppp1} with those of \Fig{ppp2}, one obtain just the phase of
$R=E+\ii B$, so its decomposition into real and imaginary parts returns
as the $E$ and $B$ polarizations, as expected from \Eq{Rkxky}.

Before closing, let us mention here one more aspect regarding polarized
emission.
Normally, Faraday rotation leads to Faraday depolarization, that is,
the cancellation of polarized emission from the line-of-sight integration
of the intrinsic polarized emission of different orientations.
Thus, in galactic magnetic field measurements with radio telescopes,
Faraday rotation was long regarded as something ``bad''.
For a helical field, however, Faraday rotation can also bring some
benefits in that it allows us to estimate the sign of magnetic helicity
and the approximate length scales of helical magnetic fields.
This property was exploited originally in the galactic context
\citep{BS14}, but it can probably also be applied to coronal magnetic
fields \citep{BAJ17}.

Studying the helicity of coronal magnetic fields further will potentially
be extremely useful in order to assess the possibility of a sign reversal
of magnetic helicity some distance above the surface of the Sun and of
other stars.
Such a sign reversal was first noted in the solar wind \citep{BSBG11},
where, far away from the Sun, the typical sign of magnetic helicity
was found to be opposite to what it is at a solar surface.
This was then later also found in simulations of simple models of global
dynamos with a conducting exterior \citep{WBM11,WBM12}.
Studying and understanding this phenomenon further will be an important
aspect of future studies.

\section{Discussion}

In this work, we have offered some speculation on how stars like the
Sun may have evolved from the times they were born to the time when they
reached an age beyond that of the present Sun.
We expect that the Sun is shortly before changing its rotation from
solar-like to antisolar-like differential rotation where the equator
rotates slower than the poles; see also \cite{Karak19} for recent
mean-field modeling of such stars.
Stars with antisolar differential rotation are also potential candidates
for displaying superflares \citep{Katsova}.
Unfortunately, there is currently no explicit observation of this
phenomenon, except for giants \citep{Kovari_etal15,Kovari_etal17}.
For main sequence stars, this idea is solely based on numerical simulations.
It must therefore be hoped that future observations can give us more
explicit evidence for this suggestion.
Helioseismic techniques might provide one such approach and has already
been partially successful \citep{Benomar+18}.
Other techniques could involve the measurement of light curves,
as has been proposed by \cite{RA15}.

We also discussed the need for a better understanding of the cyclic
nature of slowly and rapidly rotating stars.
Observations are consistent with a magnetic dipole lying in the equatorial
plane, but there is also a long-term cyclic variation that could be
compatible with the two poles alternating in their relative strengths.
But this is not yet borne out by reasonably realistic simulation data.

A promising long-term record of simulation data has been assembled by
\citep{Kapy16}.
Those simulations display a huge variety of different behaviors, including
poleward and equatorward migratory patterns, as well as short and long
cycle periods, all in one and the same run.
It would therefore be useful to produce similar data for stars with
different Rossby numbers in an attempt to better understand the various
observation signatures, including the various transitions and the proper
position of the Sun in this vast parameter space.

Finally, we turned attention to more direct inspection techniques using
linear polarization data.
This is motivated by the possibility that standard inversion techniques
to obtain the magnetic field might be severely flawed by the fact that
no safe $\pi$ disambiguation technique exists that tells us whether the
magnetic field vector points for forward or backward.
This problem results from the fact that polarization ``vectors'' are
not proper vectors as they do not have neither head nor tail.
In fact, just to obtain the sign of magnetic helicity, it is, under some
conditions, not even necessary to disambiguate the polarization vector.
The sign of handedness can in fact be obtained directly from the  linear
polarization---without resorting to the magnetic field.

An intermediate approach here is to first determine the magnetic
field, but then to make it ambiguous again by estimating $Q+\ii U$
from $(b_x+\ii b_y)^2$.
This sounds somewhat odd, but it has the advantage that one does then
not need to worry about wavelength dependencies of the line spectra of
Stokes $Q$ and $U$.
This has been discussed in detail in the work of A.\ Prabhu
(private communication).

A different approach to using Stokes $Q$ and $U$ directly is in connection
with the determination of magnetic helicity in solar and stellar coronae.
This idea was originally proposed for edge-on galaxies \citep{BS14},
but it can equally well if he applied to the Sun and other stores.
The work of \cite{BAJ17} suggests, however, that one may have a better
chance of exploiting this technique by using millimeter wavelengths
rather than infrared.
In any case, it is necessary to measure polarized intensity over a broad
range of different wave lengths.
This has now become possible with the emergence of more refined detector
technology.
In the context of galactic polarization measurements, this technological
advance is what led to the development of Faraday tomography \citep{BB05}.
This is based on only work of \cite{Bur66}, who recognized that the
line-of-sight integral of linear polarization is the same as a Fourier
integral and can therefore be inverted, provided one covers a sufficiently
broad range of wavelengths.
This was exactly the problem that was difficult to overcome in the early
days of radio astronomy, where observations could only be carried out
in a small number of frequency bands

A reliable measurement of magnetic helicity in the solar corona is greatly
helped by the possibility to use the moon as a perfect coronograph.
The total eclipse during the IAU symposium has not yet been utilized
for that purpose, but this would hopefully change in the near future.

\begin{acknowledgements}
This work was supported by the National Science Foundation
under the grant AAG-1615100 and the Swedish Research Council
under the grant 2019-04234.
We acknowledge the allocation of computing resources provided by the
Swedish National Allocations Committee at the Center for Parallel
Computers at the Royal Institute of Technology in Stockholm.
\end{acknowledgements}


\begin{discussion}

\discuss{Klaus Strassmeier}{
It is indeed true that we have found no good evidence for anti-solar
differential rotation in main sequence stars, but only in giants
and subgiants.
The stars with anti-solar differential rotation were not solar-like
stars when on the main sequence, but may have been some sort of Ap stars without
a sign of an outer convection zone.
Do we see differential rotation at all in Ap stars?
Or, if seen in solar-like main sequence stars, what process switches
differential rotation from solar- to anti-solar?
}
\discuss{Axel Brandenburg}{
The process causing this switching from solar-like to anti-solar
differential rotation is an emerging dominance of meridional circulation
over the Reynolds stress.
Subgiants could also have meridional circulation in a thin outer
convective layer causing anti-solar differential rotation.
}

\discuss{Christopher Karoff}{
What would happen to the topography of the magnetic field as the stars
change to anti-solar differential rotation?
}
\discuss{Axel Brandenburg}{
According to the simulations, it should be the same, so the magnetic
field would still be poleward migrating.
This is because in the solar-like differential rotation regime, it has
been very difficult to reproduce solar-like equatorward migration.
Therefore, poleward migration has been obtained in both cases.
In reality, however, this might not be true, and so poleward migration
might emerge only with the moment that the star begins to display
antisolar differential rotation.
}

\discuss{Moira Jardine}{
One of the ways to confuse a periodogram is to have spots whose lifetime
is less than the rotation period of the star.
Can you say something about any trends in spot lifetime with accuracy?
}
\discuss{Axel Brandenburg}{
Sunspots are known to have a large range of lifetimes from half a day to
three months.
Their decay times scale with their surface area, so for the Sun,
the larger and more dominant spots do have lifetimes longer than the
rotation period.
}

\end{discussion}
\end{document}